    \documentclass[5p,twocolumn]{elsarticle}
     \usepackage{tabularx} 
 
    \usepackage[breakable,skins]{tcolorbox}
    \usepackage{xcolor}
    \usepackage{hyperref}
    \hypersetup{
        colorlinks=true,
        linkcolor=black,
        filecolor=magenta,      
        urlcolor=cyan, 
        }
    \usepackage{graphicx,cuted} 
    \usepackage{bbding} 
    \usepackage{amssymb,latexsym,amsfonts,amsmath,amsthm}
    \usepackage{diagrams}
    \usepackage{arydshln}
    \usepackage{verbatim} 
     \usepackage{diagbox}\usepackage{float}
    \usepackage{temporal}
    \usepackage{comment}
    \usepackage{booktabs} 
    \usepackage{amssymb,latexsym,amsfonts,amsmath,amsthm}
    \usepackage{url}
    \usepackage{xurl}
    \usepackage{color}
    \usepackage{subfigure}
    \usepackage{flushend}
    \usepackage{enumerate}
    \usepackage[normalem]{ulem}
    
    \usepackage{array}
    \usepackage{makecell}
    \usepackage{caption}
    
    \def\change#1{{{#1}}}

    \usepackage{amssymb}
    \usepackage{enumitem}
    \setlist[enumerate,1]{label=\arabic*,leftmargin=2.5mm}
    \setlist[enumerate,2]{label=\alph*),leftmargin=4.5mm}

    \newtcolorbox{fancyquotes}{%
        enhanced jigsaw, 
        breakable,      
        frame hidden,   
        left=0.5cm,       
        right=0.5cm,  
                    parbox=false,
    }
    \usepackage{natbib}
      \setcitestyle{authoryear,open={(},close={)}}
    \journal{Annual Reviews in Control}
    
\begin{document}
    
    \begin{frontmatter}
    
    \title{{\bf Formal Synthesis of Controllers for Safety-Critical Autonomous Systems:   Developments  and Challenges}}

    \author[sjtu]{Xiang Yin}
    \ead{yinxiang@sjtu.edu.cn}  
    \author[tju]{Bingzhao Gao}
    \ead{gaobz@tongji.edu.cn}
    \author[xmu]{Xiao Yu}
    \ead{xiaoyu@xmu.edu.cn}
    
    \address[sjtu]{Department of Automation, Shanghai Jiao Tong
    University, Shanghai 200240, China}
   
    \address[tju]{School of Automotive Studies, Tongji University, Shanghai 201804, China} 
        
     \address[xmu]{Institute of Artificial Intelligence, Xiamen
    University, Xiamen 361005, China} 

    \begin{abstract}
\change{
In recent years, \emph{formal methods} have been extensively used in the design of autonomous systems. By employing mathematically rigorous techniques, formal methods can provide fully automated reasoning processes with provable safety guarantees for complex dynamic systems with intricate interactions between continuous dynamics and discrete logics.  This paper provides a comprehensive review of formal controller synthesis techniques for safety-critical autonomous systems. Specifically, we categorize the formal control synthesis problem based on diverse system models, encompassing deterministic, non-deterministic, and stochastic, and various formal safety-critical specifications involving logic, real-time, and real-valued domains. The review covers fundamental formal control synthesis techniques, including abstraction-based approaches and abstraction-free methods. We explore the integration of data-driven synthesis approaches in formal control synthesis. 
Furthermore,  we review formal techniques tailored for multi-agent systems (MAS), with a specific focus on various approaches to address the scalability challenges in large-scale systems. Finally, we discuss some recent trends and highlight research challenges in this area.} 
    \end{abstract}
    
    
    \end{frontmatter}


    \section{Introduction}
    With the rapid advancement of information technology, control technology and artificial intelligence, \emph{autonomous systems} are now extensively utilized in various areas of our society, including industrial manufacturing systems \cite{theorin2017event,fragapane2022increasing}, intelligent transportation systems \cite{schwarting2018planning,chen2022milestones}, and healthcare in daily life \cite{attanasio2021autonomy,navarro2021proximity}.  Notably, these autonomous systems often operate within complex and dynamic environments with uncertainties and potential adversarial components. A defining characteristic of autonomous systems is their comprehensive integration of perception and decision-making modules, accompanied by substantial computational capabilities. Consequently, a central challenge in autonomous systems lies in how to effectively make decisions online in order to effectively leverage information acquisition and computational resources, ensuring accurate and timely commands in the face of the complexities of their operational environments.
    
    A fundamental characteristic of autonomous systems is that they are typically classified as \emph{safety-critical systems} \cite{wolf2017safety,xiao2023safe,dawson2023safe}, which implies that any malfunction in their functionalities could result in significant catastrophes. For instance, in autonomous vehicles, safety is consistently the foremost concern, as any error in their control software has the potential to endanger the lives of passengers \cite{zhao2022formal,selvaraj2022formal}. Similarly, in manufacturing systems, production robots must operate in a correct sequence to prevent collisions or buffer overflows \cite{chryssolouris2013manufacturing}. 
    Another example is scenario of urban search and rescue robots, where they need to  ensure both the safety of the robot itself and the individuals being rescued; this requires to  navigate to a desired region promptly while avoiding unknown obstacles \cite{delmerico2019current}.

    Nevertheless, synthesizing controllers for autonomous systems poses a significant challenge due to the complex dynamics of the systems and the intricate design objectives involved. In particular, autonomous systems are also characterized as \emph{cyber-physical systems}, featuring continuous dynamics at the physical level and embedded control logics at the software level \cite{alur2015principles,lee2016introduction,platzer2018logical}. 
    \change{Consequently, ensuring safety for this class of hybrid dynamic systems     necessitates addressing design concerns at both the low-level, such as collision avoidance, and the high-level, ensuring logical correctness so that tasks are executed in the correct order.}

    \change{The hybrid nature of autonomous systems and the safety specifications not only make control strategies designed susceptible to errors but also may contribute to a prolonged and time-consuming design phase.
    Designing autonomous systems based solely on human experience may not be sufficient to address the inherent complexity challenges.
    To ensure safety and streamline the design process, one may seek that control synthesis procedures possess the following two key properties:}  
    \begin{itemize}
    \item 
    Fully Automated: The control synthesis procedure should be fully automated, eliminating the need for designers to investigate design objectives on a case-by-case manner. Instead, the designer can formalize the specifications and the underlying system as structural inputs. Subsequently, the synthesizer can \emph{automatically} generate the control law or even control codes without requiring human intervention.
    \item 
    Correctness Guarantees: Upon the generation of the control law by the synthesizer, \emph{formal guarantees} regarding its correctness can be deduced through rigorous mathematical tools. This eliminates the need for manual examination and redesign of controllers, thereby not only reducing the duration of the design phase but also enhancing the reliability of the controller.
    \end{itemize}

\change{Generally speaking, formal methods refer to the use of mathematically rigorous techniques to ensure the correctness of complex systems.}
It was  originally developed within the field of computer science in the context of software verification and synthesis, offer a powerful framework to fulfill the requirements aforementioned \cite{clarke1996formal,manna2012temporal}. Notably, by leveraging rigorous mathematical techniques to formalize both the system model $\Sigma$ and the desired specifications $\varphi$, one can ensure provably correct guarantees for reasoning processes and results. Within the context of formal methods, two fundamental problems emerge: \emph{verification} problems and \emph{synthesis} problems. In the verification problem, the task is to check whether a given (closed-loop) system model satisfies some specific desired specifications. Conversely, in the synthesis problem, the system model is open and reactive, and the objective is to synthesize a program to determine inputs online so that the given specification can be enforced. 
    The formal synthesis problem is inherently related to the control problem, since the program is essentially a feedback controller.  
    This design paradigm of formal synthesis is also commonly referred to as ``\emph{correct-by-construction}" control synthesis.
    
    Over the past decades, there has been a significant increase in attention towards applying formal methods to control synthesis of autonomous systems \cite{belta2017formal,kress2018synthesis,luckcuck2019formal,belta2019formal,li2019formal,mehdipour2023formal}. A major catalyst for this trend is the capability of formal methods to provide a unified approach for describing and managing heterogeneous desired safety-critical design objectives, encompassing both low-level dynamics and high-level logics. For instance, in autonomous robots, a fundamental task involves the ``reach-avoid task" \cite{summers2010verification,fridovich2020confidence} requiring that the robot should visit a target region $\textsf{Goal}$ while avoiding unsafe regions $\textsf{Bad}$ during its operations. This task can be succinctly captured by a linear temporal logic (LTL) formula \cite{baier2008principles}, such as $\varphi=\mathbf{F} \textsf{Goal} \wedge \mathbf{G} \neg \textsf{Bad}$, where $\mathbf{F}$ denotes ``eventually" and $\mathbf{G} $ denotes ``always". Moreover, LTL specifications empower users to articulate more intricate safety-critical requirements for autonomous systems, such as surveilling a target region infinitely often or executing various sub-tasks in a specified order.

    This paper aims to offer a thorough survey of the basic methods, recent advances, and the state-of-the-art in the application of formal control synthesis techniques to safety-critical engineering cyber-physical systems, with a particular emphasis on autonomous systems such as teams of autonomous robots. 
    Particularly, in this paper, we adopt a unified approach by \emph{treating safety requirements as the satisfaction of a logical formula  $\varphi$}. 
    As discussed earlier, this formulation offers a unified method that bridges the gap between the satisfaction of physical constraints and the correctness of high-level logics. 
    In summary, the primary focus is on addressing the formal control synthesis problems of various forms within the following context. 
    \begin{fancyquotes}
    \emph{Given an autonomous system $\Sigma$ operating in an uncertain environment and a desired specification formula $\varphi$, synthesize a feedback control $C$ such that the closed-loop system $\Sigma_C$ satisfies the specification $\varphi$ with a certain level of formal guarantees. }
    \end{fancyquotes}

\begin{figure*}[t]
      \centering
      \includegraphics[width=0.95\textwidth]{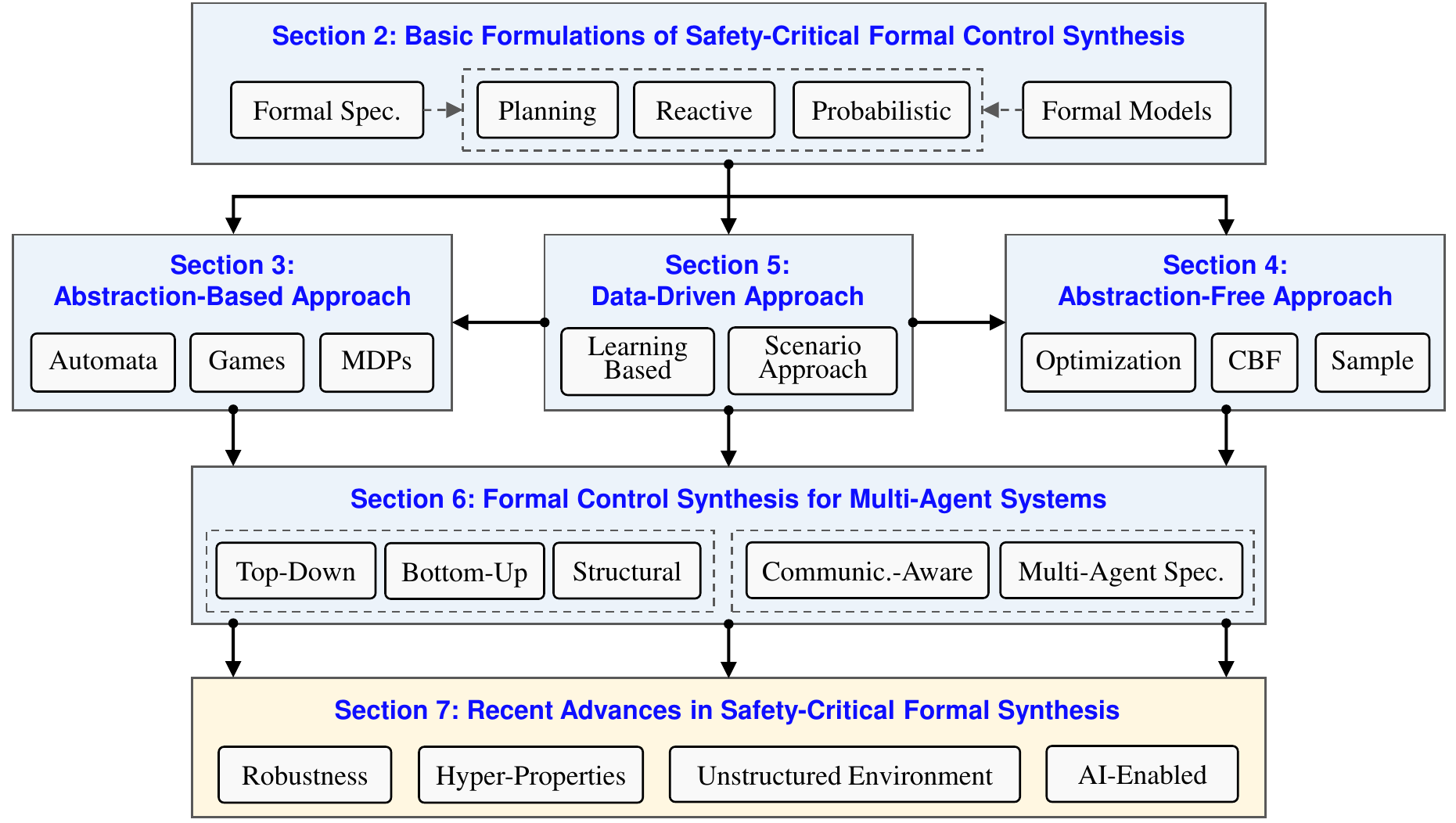}
      \caption{Organization of the paper.}\label{fig:structure}\vspace{5pt}
    \end{figure*}

    The structure of this paper is shown in Figure~\ref{fig:structure}. 
    In Section~\ref{sec:2}, we first classify formal control synthesis problems for autonomous systems. We also introduce some basic specification languages widely used in safety-critical formal control synthesis. 
    In Section~\ref{sec:3}, we discuss how to synthesize controllers when systems are abstracted as symbolic models. Section~\ref{sec:4} focuses on how to synthesize controllers directly based on the continuous dynamics of the system without discretizing the state spaces. When the system model is unknown to the user, data-driven synthesis approaches are introduced in Section~\ref{sec:5}. Section~\ref{sec:6} extends control synthesis techniques to the case of multi-agent systems (MAS), where local agents may be coupled with each other by networks. We particularly focus on how to synthesize controllers for large-scale systems in a scalable manner and how to address new safety-critical requirements that arise in multi-agent systems. 
    Finally, we conclude the paper in Section~\ref{sec:7} by discussing more recent trends and research challenges formal control synthesis for autonomous systems.

    \section{Formal Control Synthesis Problems}\label{sec:2}
    In this section, we categorize the formal control synthesis problems considered in this paper. Specifically, based on whether the system model is deterministic, non-deterministic, or probabilistic, the underlying synthesis problems are classified as planning problems, reactive synthesis problems, and probabilistic synthesis problems. Additionally, we review some commonly used formal specification languages for autonomous systems. 
    
    \subsection{Formal System Models}
    In this works, we follow the formalism by explicitly considering the dynamic of the autonomous system in an environment as a \emph{plant model}.  Roughly speaking, a system model is a tuple $\Sigma=(X,U,f)$,  where $X$ is the state space, $U$ is the action space, and $f$ is the transition relation \cite{tabuada2009verification}.  
    In general, $X$ and $U$  can be uncountable infinite sets such as real numbers when the system is continuous.  
    When they are countable sets, the system $\Sigma$ is referred to as a \emph{symbolic system}. 
    In numerous applications, a system $\Sigma$ is also augmented with a labeling function that specifies the atomic properties holding at each system state. 
    Then, depending on the structure of the transition relation, 
    formal models can be further categorized as follows. 
    
    \textbf{Deterministic Models: } 
    A system is categorized as deterministic when the transition relation conforms to the form $f: X\times U\to X$. In essence, such a system executes control commands perfectly without encountering any disturbances. Consequently, for these ideal systems, a feedback control strategy is unnecessary, and one can simply design an \emph{open-loop plan} for execution.
    For continuous systems, it can be the standard linear or nonlinear systems without disturbances. 
    For symbolic systems, deterministic finite state automata or labeled transition systems represent the most conventional models. In the cases where the system exhibits countable but infinite states, Petri nets \cite{murata1989petri,reisig2012petri,giua2018petri}  or pushdown automata \cite{griffin2008note,schmuck2016supervisory} offer suitable tools for capturing  unbounded state spaces.
    
    \textbf{Non-Deterministic Models: }
    A system is categorized as non-deterministic when the transition relation conforms to the form $f: X\times U\to 2^X$.  
    For continuous systems, this can be differential inclusion systems with set-value mappings \cite{aubin2012differential}. In the context of symbolic systems, non-deterministic transition systems or finite-state automata with uncontrollable events are common representations \cite{cassandras2008introduction}.
    This modeling approach is typically employed to represent environments characterized by uncertainty, disturbances, or adversarial elements, particularly in a zero-sum context. In this non-deterministic setting, for a given pair of state and control input, the system may move non-deterministically to a set of possible states; which specific state to go depends on the choice of the environment.  
    
    Note that the physical interpretation of transition non-determinism varies depending on the specific problem scenarios. In a basic case, non-deterministic transitions may capture operation in a perturbable environment, where the actuator cannot execute a command perfectly due to unmodeled uncertainty. Additionally, non-deterministic transitions can represent the presence of an uncontrollable agent; for instance, a robot operating in a workspace with freely moving passengers. Moreover, non-determinism can capture unknown environments, where a state may have multiple possible properties before the agent actually visits it. The mathematical model of non-deterministic systems provides a unified framework to represent these diverse physical scenarios.

    \textbf{Stochastic Models: } 
    It is worth noting that, in non-deterministic models, there is no a priori knowledge regarding the likelihood of each possible transition. By incorporating this probabilistic priori information into the model, we obtain a stochastic model characterized by a transition relation  of the form $f: X\times U\to \textsf{dis}(X)$, where $\textsf{dis}(X)$ is a probabilistic distribution on the state space $X$. This model is commonly denoted as the Markov decision process \cite{puterman2014markov,kumar2015stochastic}. 
    Note that, for both non-deterministic or stochastic models, feedback control strategies are strictly necessary as the controller needs to adjust its plan online based on the behavior of the environment. 
     
    \textbf{Multi-Player Models: } 
    The three classes of models aforementioned are widely used in capturing the robust operation of a single (team of) system in environments with uncertainties. Note that, although non-deterministic transitions are allowed in these models, the environment itself lacks a distinct objective. In certain scenarios, an additional player, distinct from the controlled system, may be introduced, each with its own specific objective. This objective may not be entirely adversarial to that of the controller. 
    For instance, the input set can be partitioned as $U=U_1\dot{\cup} U_2$, and the transition function can be formulated as $f: X\times U_1\times U_2\to X$ for the case concurrent decision-making \cite{de2007concurrent} or $f: \cup_i(X\times U_i)\to X$ for the case turn-based decision-making \cite{brenguier2016non}.
    Stochasticity can also be integrated into multi-player models, leading to stochastic games \cite{kvretinsky2022comparison,zhang2021multi}.
    
    Throughout this paper, our focus will be on reviewing the formal synthesis for the first three classes of systems. Specifically, we assume that all decision-makers are cooperative against the uncertain environment sharing the same overall objective. Multi-player decision-making in the context of non-zero-sum games is beyond the scope of this survey.
    
    \subsection{Formal Specification Languages}
    
    Given a system model $\Sigma$, whether in an open-loop configuration or closed-loop under control, each individual execution of the system generates a (finite or infinite) sequence of states (with labels if needed), which is referred to as a \emph{trace}. The aggregate of all possible traces generated by $\Sigma$ is called its \emph{language}, denoted as $L(\Sigma)$. Essentially, formal specifications serve the purpose of characterizing whether the language of $\Sigma$  is  a ``good" language.

    It is important to note that, in general, a specification needs to evaluate $L(\Sigma)$ as a whole rather than individually evaluating each trace in it. 
    A particular but notably powerful case is that of \emph{linear-time specifications}, where the satisfaction of the specification can be evaluated by inspecting the satisfaction of each individual trace. For instance, ``reach-avoid" stands as the most widely used linear-time specification. In the domain of safety critical control of autonomous systems, the following linear-time specifications are widely employed.
    
    Linear Temporal Logic (LTL) is one of the most widely used formal specifications for describing the safety and correctness of autonomous systems \cite{pnueli1977temporal}.  Specifically, an LTL formula $\varphi$ is evaluated over infinite traces. In addition to standard Boolean operators, LTL allows temporal operators such that ``always" $\mathbf{G}$, ``next" $\mathbf{X}$, ``eventually" $\mathbf{F}$ and ``until" $\mathbf{U}$.  
    Therefore, it supports complex safety-critical requirements such that ``\emph{should not do task $B$ before task $A$ is finished}". 
    A notable feature of LTL is that,  for an arbitrary LTL formula $\varphi$, it can be accepted by a non-deterministic B\"{u}chi automaton or a deterministic Rabin/parity automaton.   
    Therefore, automata-theoretical approaches can be applied to model-checking and control synthesis problems. 
    There are also many useful variants of LTL formuale such as co-safe LTL (scLTL) \cite{kupferman2001model}, LTL over finite traces (LTLf) \cite{de2013linear} and Truncated LTL (TLTL) \cite{li2017reinforcement}. 
    Notably, scLTL formulae are widely used in describing tasks that can be satisfied within a finite horizon. 
    Specifically, scLTL requires that negation operator can only be applied in front of atomic proposition; therefore, temporal operator ``always" is not allowed. 
    The main property of scLTL is that any infinite trace satisfying the formula has a ``good" prefix and can be accepted by deterministic finite-state automata. 
    
    Note that the applications of LTL are restricted to traces that are purely logical and contain no real-time information. However, in real-world autonomous systems, the traces generated by the systems consist of real-valued signals in the time domain. Therefore, it becomes imperative to further extend LTL to encompass both continuous state space and real-time considerations. Metric Temporal Logic (MTL) stands out as a prominent extension of LTL by introducing constraints on temporal operators with intervals of real numbers \cite{koymans1990specifying,thati2005monitoring}. For instance, $\mathbf{F}_{[4,5]}\phi$ denotes that the task $\phi$ can be satisfied within 4 to 5 time units from the present moment. 
    Signal Temporal Logic (STL) further extends MTL to traces over real-valued state spaces \cite{maler2004monitoring}. Specifically, by considering a set of atomic predicates $\mu(x)$ that are evaluated over real-valued signals, STL allows for checking the satisfaction of a spatial signal in real-time. \change{Furthermore, by introducing the robust semantics of STL \cite{fainekos2009robustness,donze2010robust,deshmukh2017robust,zhong2021extending}, one can quantitatively evaluate the satisfaction degree of the task rather than receiving a Boolean answer.}

    The previously mentioned LTL, MTL, and STL, along with their variants, all belong to the category of linear-time properties. However, many applications require \emph{branching-time} properties to capture the desired specifications. For example, one might necessitate that at each instant, the robot has an exit path to a safe region, but it is not necessarily obliged to execute it. This requirement cannot be captured by linear-time specifications but can be expressed using CTL$^\star$ with temporal operators and path quantifiers, which is an extension of Computation Tree Logic (CTL) \cite{emerson1986sometimes}. In the context of supervisory control of discrete-event systems, the well-adopted non-blockingness condition \cite{wonham2019supervisory,yin2015synthesis,yin2018synthesis} essentially represents a branching-time requirement. This condition can be captured by a CTL$^\star$ formula such as $\textbf{A}\textbf{G}\textbf{E}\textbf{F}\textsf{Goal}$ \cite{ehlers2017supervisory}, 
    where $\textbf{A}$ and $\textbf{E}$ are path quantifiers ``for all" and ``exists", respectively.

    \subsection{Formal Control Synthesis Problems}
    As discussed, the control synthesis problem and the corresponding formal guarantees vary for different types of system models. Here, we categorize the formal control synthesis problem into the following three categories.

    \textbf{Planning Problem: }
    For a deterministic system model, employing a feedback control strategy is unnecessary.
     Instead, it suffices to identify an open-loop input sequence under which a \emph{path} $\rho$ with finite or infinite states (or state labels) can be generated by system $\Sigma$. For planning problems, the consideration involves a linear-time property specification $\varphi$, and the corresponding formal guarantee is expressed as $\rho\models \varphi$.

    \textbf{Reactive Control Synthesis Problem}
    For a non-deterministic system model $\Sigma$, the controller $C$ is a feedback strategy that issues the next control input based on the observation history. In general, memory is required to realize the controller $C$, and the resulting closed-loop system $\Sigma_C$ can be considered as a synchronized product between the system model and the controller model. In this case, the trace generated by the system is not a single one but a language, and the specific trace generated online depends on the choice of the non-deterministic environments. The corresponding formal guarantee is expressed as $\Sigma_C \models \varphi$, meaning that the language generated by $\Sigma_C$ satisfies the specification. Particularly, when $\varphi$ is a linear-time specification, the formal guarantee can be equivalently expressed as $\forall \rho \in L(\Sigma_C):\rho\models \varphi$. 
    This is essentially a robust condition meaning that the system is always safe no matter what the environment does.

    \textbf{Probabilistic Synthesis Problem} 
    When the system model $\Sigma$ incorporates probabilistic information in transitions, it becomes possible to discuss the likelihood of each trace generated. For instance, the system may not always satisfy the specification, but the probability of failure is very small. A typical formal guarantee for probabilistic synthesis is expressed as $\mathbb{P}(\Sigma_C \models \varphi)\geq 1-\epsilon$, where $\mathbb{P}$ is a probability measure, and its event space encompasses the set of all possible languages.
    Finally, we remark that, for all these three types of formal synthesis problems, beyond the qualitative satisfaction of formal specifications, numerous quantitative cost criteria can also be imposed to evaluate the performances of the controller, which may further lead to multi-objective synthesis problems and constrained optimization problems.


    \section{Abstraction-Based Formal Control Synthesis}\label{sec:3}
    \begin{table*}[bp] \small 
        \centering
    \begin{tabular}{ | >{\centering\arraybackslash}m{0.8cm} | >{\centering\arraybackslash}m{6cm}| >{\centering\arraybackslash}m{4.2cm} |>{\centering\arraybackslash}m{5.2cm} | } 
      \hline
            & \textbf{Planning} &   \textbf{Reactive Control} & \textbf{Probabilistic Synthesis} \\ 
      \hline
      \textbf{LTL} 
      & 
      {
      \scriptsize
      \cite{crouse2023formally,yu2022security,yang2020secure,kalluraya2023resilient,huang2022failure,zhao2023explore,yu2021distributed,wolff2013efficient,wolff2016optimal,vasile2020reactive,vasile2013sampling,ulusoy2014receding,tabuada2003model,schillinger2018simultaneous,lv2023optimal,luo2021abstraction,liu2023nngtl,li2023temporal,li2023fast,kurtz2023temporal,kloetzer2020path,kloetzer2016multi,kloetzer2008fully,kantaros2022perception,kantaros2020reactive,kantaros2019temporal,kantaros2020stylus,he2015towards,guo2018multirobot,guo2015multi,gol2015temporal,gol2014additive,gol2013language,fang2022decentralized,diaz2015correct,bhatia2010sampling,banks2020multi,aydin2013temporal,gol2015temporal,diaz2015correct, kloetzer2016multi,vasile2017minimum,kloetzer2020path,grover2021semantic,cohen2021model,hustiu2021optimal, kamale2021automata, liang2022fair, zhao2023explore,huang2023synthesis,smith2011optimal,ulusoy2013optimality}
      }
      & 
      {
      \scriptsize
      \cite{wang2023conformal,zhou2023vision,zhou2023local,niu2020optimal,kalluraya2023resilient,huang2022failure,zhao2023explore,yu2021distributed,vasile2020reactive,vasile2017minimum,schuppe2020multi,ramasubramanian2020secure,maniatopoulos2016reactive,li2023temporal,li2021reactive,kulkarni2018compositional,kantaros2022perception,kantaros2020reactive,huang2023synthesis,hashimoto2022collaborative,guo2018multirobot,grover2021semantic,gol2015temporal,fu2016synthesis,diaz2015correct,alur2018compositional,ulusoy2013temporal, liu2013synthesis, wolff2013efficient, lignos2015provably, schmuck2017dynamic, balkan2017mode, peterson2020decentralized, keroglou2020communication,sakakibara2020line}
      }
      & 
      {
      \scriptsize
      \cite{wang2023conformal,kalluraya2023multi,niu2020optimal,xie2021secure,wolff2013efficient,ramasubramanian2020secure,oura2020reinforcement,oh2022hierarchical,luo2021abstraction,li2019formal,lavaei2022automated,lacerda2014optimal,kazemi2020formal,kantaros2018sampling,hashimoto2022collaborative,hasanbeig2019reinforcement,hahn2019omega,haesaert2018temporal,haesaert2020robust,guo2023hierarchical,guo2018probabilistic,ghasemi2020task,fu2015synthesis,ding2014optimal,cui2023security,cai2021optimal,bozkurt2020control,alshiekh2018safe,wolff2012robust, lacerda2014optimal,ulusoy2014incremental, haesaert2018temporal, lacerda2019probabilistic, lavaei2020formal, oh2022hierarchical, kulkarni2022opportunistic, sun2022neurosymbolic, schon2022correct, hashimoto2022collaborative,chen2023entropy}
      }
      \\ 
      
    
      \hline
        \textbf{MTL}
      &   
      {
      \scriptsize
      
        \cite{kurtz2021more,li2022online,wang2022decentralized,verginis2019reconfigurable,saha2016milp,pant2017smooth,kurtz2021more,karaman2008vehicle,cardona2023temporal,karaman2008vehicle, liu2014switching, alqahtani2018mtl, andersson2017control, alqahtani2018predictive, cardona2022planning}
      }
      &
      {
      \scriptsize
        \cite{niu2020control,li2021policy,saha2016milp,pant2017smooth,liu2014switching,cardona2022planning,andersson2017control,andersson2018human} 
      }
      & 
      {
      \scriptsize
      \cite{abbas2014robustness,baharisangari2022distributed,montana2016sampling,niu2020control,fu2015computational,li2021policy} 
      }
      \\
    
      \hline
        \textbf{CTL}
       &   
      {
      \scriptsize
      NA
      }
      & 
      {
      \scriptsize
      \cite{gutierrez2017model,jiang2006supervisory,ehlers2017supervisory,kochaleema2019methodology,ehlers2014bridging,pan2015model,kupferman2000open,pnueli1989synthesis,jiang2006supervisory,benevs2020parallel} 
      }
      & 
      {
      \scriptsize
      \cite{lahijanian2010motion,yoo2013provably,vcevska2019counterexample,pajiccontext,baier1997symbolic,wu2016counterexample,zhang2015learning,cizelj2013negotiating,gerasimou2015search,wu2015counterexample,kamide2021inconsistency,qian2020safe,lahijanian2015formal,puggelli2014robust,lahijanian2011temporal,nikou2017probabilistic,wagner2011finite,huth2005finite,baltazar2007exogenous,baier2004controller} 
      }
      \\ 
      \hline
    \end{tabular}
        \caption{Summary of works on  formal control synthesis with symbolic models.}
        \label{tab:my_label222} 
    \end{table*}

    In this section, we focus on discussing formal control synthesis techniques based on symbolic system models.  Although symbolic models are inherently limited by discrete state spaces, techniques associated with symbolic systems continue to play fundamental roles in formal control synthesis for several reasons. 
    \begin{itemize}
      \item  
        First, a considerable number of safety-critical systems inherently exhibit symbolic characteristics and it is sufficient enough to consider symbolic models at the task planning level.
        A notable example of this case is the autonomous robot with embedded controllers, where the control software operates within a symbolic framework \cite{kress2018synthesis}. 
        Another example is the manufacturing process, where discrete states represents the configurations of the machines and complex sequences of events are  involved to drive the evolution of configurations \cite{theorin2017event}.  
      \item 
      Second, although many more complex systems involve continuous state-spaces and time-driven dynamics, they can be \emph{abstracted} as symbolic system, either precisely or approximately.  
      This issue will be discussed in this section. 
      Therefore, synthesis techniques can still be applied to their abstractions, and the resulting symbolic controllers can be refined back to control the original systems \cite{tabuada2009verification}.   
      \item
       Finally, it should be noted that most of the widely used formal specifications can be expressed through symbolic models. 
       For instance, regular languages find representation through finite state automata, while linear temporal logic specifications can be captured by nondeterministic B\"{u}chi automata. Consequently, symbolic models offer a unified framework for aligning the dynamic behavior of a system with the  task processes under investigation.  
    \end{itemize}
    Therefore, a comprehensive understanding of synthesis techniques for symbolic systems becomes the foundational step towards formal synthesis for more complex systems. 
    
    \subsection{Construct  Symbolic Models via Formal Abstractions}
    As mentioned earlier, the initial phase of abstraction-based formal synthesis involves constructing the symbolic model from the original dynamic system through abstraction. Various approaches exist for building formal abstractions, and the choice depends on the nature of the underlying dynamic systems. Here, we briefly outline some commonly used techniques.
    
    The most straightforward approach is to partition the operational state space based on atomic properties of interest. In this method, connected regions with the same properties can be abstracted as a single symbolic state. This technique is typically employed in broad and non-complex environments, where the dynamics of the agent are not a primary concern. In such scenarios, feasible plans at the physical level can readily be found to navigate the agent from one state to another. 
    For general dynamic systems operating in polygonal environments, computationally efficient algorithms have been developed in \cite{tabuada2003model, kloetzer2008fully, fainekos2009temporal} to automatically partition the state space into polygonal triangulations. Based on these triangulations, vector fields can be constructed to generate trajectories navigating between each abstract state \cite{belta2005discrete}.
    
    Another approach is to discretize the state-space so such some formal relationship between the abstract system and the concrete system can be established.   For example, for planning problems for deterministic systems, one can use the notion of (approximate) (bi-)simulation relations \cite{girard2007approximation,pola2008approximately,girard2009approximately,zamani2011symbolic}. For reactive control synthesis problems, the alternating simulation relation can be used to handle transition non-determinism for each control action \cite{alur1998alternating,pola2009symbolic,hou2019abstraction}.  
    These relations have also been extended to the probabilistic setting \cite{zhong2023automata,zamani2014symbolic,zamani2015symbolic,lavaei2022automated}.    There are also many other approaches to construct formal absractions, such as $l$-complete approximations \cite{moor2002discrete,schmuck2014asynchronous} and feedback refinement relations \cite{reissig2016feedback,ren2020symbolic,khaled2022framework}.

    \subsection{Methods for Path Planning}
    For deterministic symbolic systems, the path planning problem aims to find a feasible sequence such that a given specification $\varphi$ is satisfied.  
    The primary method for addressing this type of problem is through an automata-theoretical approach \cite{vardi2005automata}. 
    Notably, for any given specification $\varphi$, a satisfiable path in the plant represents an instance of violation of the formula $\neg\varphi$. Consequently, by conducting model checking on the negation of the task over the plan model, a feasible plan can be readily identified. 
    Essentially, this approach involves first establishing the synchronization product  of the specification automaton and the plant model.
    
    The automata-theoretical approach has been employed by most formal planning algorithms, either explicitly or implicitly. 
    The basic approach involves building the entire product system all at once and then performing a graph search for feasible and optimal paths over the product space, depending on the accepting condition of the specification automaton and the numerical metric for optimality \cite{smith2011optimal, ulusoy2013optimality,guo2015multi}. However, this approach may not scale well when the size of the plant model increases. To address this, many works use the strategy of incrementally constructing the product space online as the system evolves. A very popular planning algorithm in this category is the sampling-based approach such as Rapidly Exploring Random Tree (RRT). In this approach, a tree over the product space is constructed incrementally from the initial root until a vertex formulating a satisfaction instant is visited, 
    and it turns out that this approach scales very well for systems with very large state-spaces  \cite{kantaros2018sampling,kantaros2020stylus,vasile2020reactive,liu2023nngtl}. Note that the incremental planning approach not only provides computational efficiency, but can also be applied to online path planning problems where the entire system is unknown a priori.
    \change{In the context of LTL path planning, another efficient approach is to fix the planning horizon a priori and to encode the planning problem as a  mixed-integer linear program \cite{sahin2019multirobot}.}

    \subsection{Methods for Reactive Control Synthesis}
    For non-determinsitic systems,  controllers need to react to the behavior of the environment dynamically online.  
    The exploration of formal control synthesis   commenced in the   1980s, with contributions from both the computer science and control engineering communities in the contexts of reactive synthesis for software \cite{clarke1981design,pnueli1989synthesis} and supervisory control of discrete event systems \cite{ramadge1987supervisory,ramadge1989control,wonham2019supervisory}, respectively. In the context of reactive synthesis, researchers mainly focus on \emph{open systems} operating within reactive environments.  The primary objective was to ensure that the generated output responded accurately to environmental input, thereby satisfying specified properties in the overall input/output (I/O) sequence. 
    In this setting, the possible behavior of the environment, as well as the dynamic of the agent, can be encoded as an environment formula $\varphi_{env}$ and the overall synthesis specification is of the form $\varphi_{env}\to \varphi_{task}$ \cite{kress2009temporal,liu2013synthesis,maniatopoulos2016reactive,majumdar2019environmentally}. 
    On the other hand,  in the context of supervisory control theory, an explicit plant model under control is considered with the behavior of environments being modeled by uncontrollable events.  In this scenario, a supervisor was introduced to disable the occurrence of controllable events, thereby enforcing the desired requirements.  Essentially, both of these settings address similar problems and can be approached with a more unified perspective in the context of zero-sum two-player games \cite{gradel2003automata} with qualitative or quantitative specifications.

    Note that a key feature of controllers synthesized through two-player games is their ability to maintain satisfaction of the formal task regardless of the actions of the environments. This formal guarantee essentially requires the synthesis procedure to conduct a global search to recursively eliminate unsafe states, leading to the identification of a control invariant (also referred to as a winning region) \cite{bernet2002permissive,dallal2016synthesis}. When strict global feasibility guarantees are not necessary, or when the global plant model is time-varying or even unknown a priori, path planning algorithms can also be employed in a receding-horizon fashion to yield a reactive control \cite{kantaros2020reactive,li2021reactive,grover2021semantic,kantaros2022perception}. However, such planning-based control may be myopic in the sense that it may result in a situation where no further feasible plan exists.

    In reactive control synthesis, since the feedback controller relies on online information to make decisions, it often faces the challenge of partial observability. The formal synthesis of reactive controllers with partial-observability has been extensively studied in the context of both supervisory control of DES \cite{yin2015uniform,ji2021optimal} and games with imperfect information \cite{chatterjee2007algorithms,ramasubramanian2020secure,berthon2021strategy}. In this partial-observation setting, the agent may not only need to synthesize its control strategy, but may also have to jointly design its information acquisition strategy to dynamically deploy sensors on/off online \cite{fu2016synthesis,li2023temporal}. 
    Another feature of the synthesis problem compared to the planning problem is that one can further enforce branching-time specifications such as CTL specifications \cite{jiang2006supervisory}.

    \subsection{Methods for Probabilistic Control Synthesis}
    When faced with transition probability information, the probabilistic control synthesis problem arises as a combination of the standard Markov decision process for optimality criteria and reactive control synthesis for formal specifications \cite{fu2015synthesis,cai2021optimal}. 
    This naturally results in a multi-objective optimization problem, necessitating a tradeoff between optimality performance and the satisfaction probability of the task. For example, 
    \change{\cite{abate2008probabilistic} studies probabilistic reachability and safety synthesis for stochastic hybrid systems}. 
    In the context of MDPs, \cite{ding2014optimal} addresses the problem of minimizing a specific type of expected cost across all controllers while maximizing the satisfaction probability of an LTL specification. Similarly, \cite{guo2018probabilistic,guo2023hierarchical} deals with the stochastic optimization problem, imposing the constraint that the satisfaction probability of an LTL specification is greater than a given threshold. Probabilistic Computation Tree Logic (PCTL) is also widely employed as the formal specification in probabilistic synthesis for MDP \cite{lahijanian2011temporal}.  
    
    \begin{table*}[htp]\small
        \centering
    \begin{tabular}{ | >{\centering\arraybackslash}m{0.8cm} | >{\centering\arraybackslash}m{6.2cm}| >{\centering\arraybackslash}m{4cm} |>{\centering\arraybackslash}m{5.2cm} | } 
      \hline
            & \textbf{Planning} &   \textbf{Reactive Control} & \textbf{Probabilistic Synthesis} \\ 
      \hline
      \textbf{LTL} 
      & 
      { 
      \scriptsize
      \cite{oh2020chance,kress2009temporal,karlsson2018multi,fainekos2009temporal,cai2023overcoming,gol2013language,aydin2013temporal, gol2014additive, gol2014finite, cho2017cost, karlsson2018multi,wolff2014optimization, wolff2016optimal,kloetzer2007temporal, meyer2019hierarchical, srinivasan2020control,yu2021distributed}
      }
      & 
      {
      \scriptsize 
      \cite{cai2023learning,kress2009temporal,karlsson2018multi,hasanbeig2020deep,aydin2013temporal,wongpiromsarn2012receding}
      }
      & 
      {
      \scriptsize
      \cite{oh2020chance,li2017reinforcement,hasanbeig2020deep,cohen2021model,cai2023overcoming,cai2021modular,horowitz2014compositional,oh2020chance,jagtap2020formal,ghasemi2020task, haesaert2020robust, jiang2022safe, van2023syscore,van2021multi}
      }
      \\ 
    
      \hline
        \textbf{MTL}
      &   
      {
      \scriptsize
      \cite{koymans1990specifying,donze2010robust,alqahtani2018predictive,alqahtani2018mtl,saha2017task,andersson2018human,verginis2019reconfigurable}
      }
      & 
      {
      \scriptsize
      \cite{hoxha2018mining,maler2004monitoring,koymans1990specifying,barbosa2019integrated,donze2010robust}
      }
      & 
      {
      \scriptsize
      \cite{xu2019controller,xu2018coordinated,barbosa2019integrated,xu2020differentially,xu2021provably,xu2020differentially,xu2023controller}
      }
      \\ 
    
      \hline
        \textbf{STL}
       &   
       {
      \scriptsize
      \cite{tian2022two,yu2023efficient,rodionova2022combined,rodionova2022temporal,yu2023model,yu2021hierarchical,yang2024signal,yang2020continuous,xiao2021high,van2023direct,sun2022multi,sadraddini2018formal,mehdipour2019arithmetic,liu2020distributed,liu2023safe,liu2021recurrent,liu2022compositional,lindemann2020barrier,lindemann2019control,lindemann2018control,leahy2022fast,kurtz2022mixed,kurtz2020trajectory,kalagarla2021model,haghighi2019control,gundana2022event,gundana2021event,charitidou2022receding,charitidou2021signal,raman2015reactive,buyukkocak2022control,belta2019formal} 
      }
      & 
      {
      \scriptsize
      \cite{yang2023distributed,sadraddini2018formal,sadraddini2015robust,raman2014model,lindemann2020barrier,lindemann2019control,lindemann2018control,leung2022semi,kurtz2020trajectory,gundana2022event,gundana2021event,ghasemi2022decentralized,raman2015reactive,belta2019formal} 
      }
      & 
      {
      \scriptsize
      \cite{lindemann2023conformal,lindemann2022temporal,zhao2022formal,yang2023distributed,yan2019swarm,yan2022distributed,venkataraman2020tractable,kalagarla2021model,hashimoto2022stl2vec,farahani2017shrinking,farahani2018shrinking,aksaray2016q,lindemann2021reactive} 
      }
      \\ 
      \hline
    \end{tabular}
        \caption{Summary of works on formal control synthesis  with continuous models.}\vspace{-0pt} 
        \label{tab:my_label}
    \end{table*}

    \section{Abstraction-Free Formal Control Synthesis}\label{sec:4}
    The aforementioned works on formal control synthesis in the previous section are performed on symbolic system models at the abstracted level. Although working on symbolic models provides a unified way to handle heterogeneous dynamics and specification, it faces the main challenge of the curse of dimensionality. Specifically, the size of the symbolic model grows exponentially as the dimension of the system increases. 
    Furthermore, for real-valued and real-time logics such STL, one has to work directly based on the original system since these logics usually do not have automata representations. 
    In this section, we review formal control synthesis techniques that work on continuous dynamic systems directly without the discretization of state space.
     
    \subsection{Synthesis via Optimizations}
    Given a discrete-time dynamic system over a continuous state space of the form $x_{t+1}=f(x_t,u_t)$ and a specification formula $\varphi$, the objective is to find a control input $\mathbf{u}_{0:N}$ 
    such that $\mathbf{x}_{0:N}\models \varphi$, where $\mathbf{x}_{0:N}$ is the solution state trajectory under $\mathbf{u}_{0:N}$.   
    The basic approach for solving this problem is through optimizations \cite{belta2019formal}, such as Mixed Integer Linear Programs (MILP). This approach was initially proposed by \cite{raman2014model,raman2015reactive} for control synthesis for STL specifications, where the control objective is to maximize robustness. The main idea is to encode both the dynamics of the system and the STL specifications as MILP constraints, allowing existing optimization tools to be leveraged for finding a feasible solution. Beyond STL specifications, optimization-based control synthesis has also been applied to other specification logics such as MTL  \cite{saha2016milp,da2021automatic,kurtz2021more} and LTL \cite{wolff2014optimization,sahin2019multirobot}.

    The main advantage of the optimization-based approach is that it is both sound and complete. However, the main challenge is its scalability, as the complexity of solving the entire problem grows exponentially when the horizon of the problem increases.To address this, several strategies have been explored for enhancing scalability. For instance, in \cite{sadraddini2018formal,kurtz2022mixed}, the use of fewer encoding variables was proposed to reduce the size of the optimization problem. Conversely, \cite{kurtz2021more} introduced more encoding variables to achieve a tighter convex relaxation. In \cite{sun2022multi}, scalability was improved by employing piecewise linear paths between decision instants. A recent innovative approach, based on a graph of convex sets, was introduced in \cite{kurtz2023temporal} to further enhance scalability. Another direction is to employ approximations to smooth the temporal logic semantics, enabling the application of gradient-based optimization algorithms; see, e.g., \cite{pant2017smooth,mehdipour2019arithmetic,haghighi2019control,gilpin2020smooth,kurtz2020trajectory}. These approaches are significantly more scalable compared to precise encodings. However,  they are not a complete methods and cannot be applied to complex formulae.

    \subsection{Synthesis via Model  Predictive Control}
    
    When uncertainties are encounted, the system model becomes $x_{t+1}=f(x_t,u_t,w_t)$, 
    where $w_t$ represents the disturbance from the environment that the controller should account for. To design feedback control strategies, optimization-based planning can be seamlessly integrated into the framework of \emph{Model Predictive Control} (MPC). Specifically, at each instant, one solves an optimization problem to generate an open-loop plan and only applies the first control input in the sequence to the system. The controller then repeats this process in a receding-horizon fashion, ensuring that disturbances are considered within the feedback loop.
    
    MPC-based reactive control synthesis was initially developed for LTL specifications with finite abstractions \cite{wongpiromsarn2012receding,ulusoy2014receding,gol2015temporal}. In the context of control synthesis for STL specifications, the MPC-based approach was developed together with the optimization-based approach in \cite{raman2014model,raman2015reactive}. In \cite{sadraddini2015robust}, starting from a worst-case perspective, the authors proposed a robust MPC approach to ensure feasibility under uncertainties. In \cite{farahani2017shrinking,farahani2018shrinking, yang2023distributed,yang2024signal}, the shrinking horizon strategy was applied for stochastic dynamic systems under chance constraints. Similar to the optimization-based planning problem, the key bottleneck for MPC-based synthesis is still its scalability when the horizon is very long. To address this, recently, \cite{yu2023model} proposed more scalable approaches to pre-compute feasible sets in an offline fashion, reducing the online computation burden.

    \subsection{Synthesis via Control Barrier Functions}
    To address computation challenges in control synthesis for temporal logic tasks, Control Barrier Functions (CBF), originally developed for safety-critical control of dynamic systems, have recently been adopted. The core idea of CBF involves describing the safe set as a super-level set of a smooth function  $h(x)$. By enforcing the forward invariance of the safe set, safety can be guaranteed under any possible disturbances. \change{Interested readers are encouraged to explore \cite{prajna2004safety,wieland2007constructive,wongpiromsarn2015automata,ames2016control,ames2019control,xiao2019control,yang2022differentiable} for more in-depth details and recent advances on this topic including theoretical properties of CBFs,  computations of CBF-based level sets and how to find  potential CBFs.}

    In the context of temporal logic specification, the work by \cite{lindemann2018control} pioneered the application of CBF to control synthesis for STL specifications. The key step involves approximating the satisfaction region of the STL formula as a safe region. By enforcing safety within this defined safe region, STL specifications can be reliably guaranteed. This approach has garnered significant attention and has seen various extensions \cite{xiao2021high,charitidou2022receding}. In terms of system models, the approach has been extended to diverse scenarios, including multi-agent systems \cite{lindemann2019control,lindemann2020barrier}, continuous-time systems \cite{yang2020continuous}, and systems with input constraints \cite{buyukkocak2022control}. Expanding beyond STL, CBF and its variants, such as control barrier certificates, have found application in LTL specifications \cite{srinivasan2020control} and event-based STL specification \cite{gundana2021event,gundana2022event}. The primary advantage of the CBF-based approach lies in its high scalability, as it avoids combinatorial searches for binary variables. However, it is worth noting that CBF-based approaches are generally tailored to a simpler class of STL tasks, and for more complex STL tasks, this approach may   potentially yield  no solution even when a feasible one exists.

    \subsection{Synthesis via Sampling-Based Approaches}
    Another popular synthesis approach for continuous dynamics without abstraction is the sampling-based approach. Similar to its symbolic counterpart, this approach does not directly handle the system dynamics but rather takes samples over the continuous state-space according to the system dynamics. The foundational work for this approach was introduced by \cite{karaman2009sampling,karaman2012sampling}, where Rapidly Exploring Random Graphs (RRG) were utilized for deterministic mu-calculus specifications. Subsequently, \cite{vasile2013sampling,vasile2020reactive,yu2021distributed} extended this approach to handle LTL specifications. To enhance the scalability of the sampling-based approach, researchers in \cite{bhatia2010sampling,he2015towards,luo2021abstraction} employed RRT with biased sampling, achieving state-of-the-art performance. Moreover, the sampling-based approach has been adapted for stochastic settings for LTL specifications \cite{oh2020chance}.

    \section{Data-Driven Formal Control Synthesis}\label{sec:5}
    For both abstraction-based and abstraction-free approaches discussed in the previous section, having a previous dynamic model of the system is crucial for control synthesis. However, there are scenarios where either a system model is not available beforehand, or the system  dynamics are too complex to handle. In such cases, the interaction with the system through physical or simulation tests to collect data becomes essential. 
    Furthermore, data-driven approaches can also improve the efficiency and scalability of the standard model-based approaches.  
    In this section, we delve into data-driven approaches for formal synthesis with correctness guarantees.
    \subsection{Data-Driven Formal Abstractions}
    Since symbolic models play a key role in formal synthesis, an immediate question arises: can we build symbolic models directly from data so that subsequent synthesis techniques can be applied?  Recent years have witnessed a growing body of works exploring this direction. In studies such as \cite{devonport2021symbolic,lavaei2022data}, a Probably Approximately Correct (PAC) statistical framework was developed for constructing symbolic models directly from data. Here, the fidelity of the abstraction to the underlying system is guaranteed by a PAC bound. This PAC-based framework has also been extended to different classes of systems in \cite{peruffo2022data,coppola2023data}. In \cite{cubuktepe2020scenario,lavaei2022constructing}, the authors investigated how to construct interval MDPs from data, where the transition probabilities are bounded to intervals with high confidence. The key theoretical foundation behind these approaches is the \emph{scenario approach} \cite{campi2009scenario,campi2021scenario}, which establishes formal confidence bound connections between sampled data and the underlying constraints. 
    \subsection{Data-Driven Safe Controller Synthesis}
    The scenario approach employed in data-driven formal abstractions can be further leveraged for synthesizing a safe controller directly. By utilizing control barrier certificates, the safety requirement of an unknown system can be formulated as a robust convex program, which can be further relaxed into a scenario convex program. Building on this idea, \cite{salamati2021data,salamati2022safety,nejati2023formal} addressed the data-driven safety verification problem. This approach has been extended to control synthesis problems, enabling the generation of controllers with formal safety guarantees expressed in terms of confidence bounds  \cite{chen2023data,salamati2024data}.
    Finally, in addition to the scenario approach, another formal alternative for ensuring safety via data is to use direct data-driven methods without explicitly identifying the unknown model \cite{zhong2022synthesizing,van2023direct}.

    \subsection{Formal Synthesis via Reinforcement Learning}
    Another popular approach for synthesizing specification-enforcing controllers for unknown dynamical systems is through reinforcement learning (RL), a sampling-based approach based on reward signals \cite{sutton2018reinforcement}. While most existing RL algorithms focus on optimizing scalar reward signals, recent developments have extended RL to synthesize controllers for MDPs with unknown transition probabilities under formal specifications \cite{hahn2019omega}. For instance, in the context of LTL specifications for finite MDPs, model-free RL algorithms  
    have been devised to maximize the probability of satisfying formal tasks \cite{bozkurt2020control,hasanbeig2019reinforcement,oura2020reinforcement,cui2023security}. Tabular-based RL algorithms for finite MDPs have also been developed under STL specifications \change{using both model-free algorithms  \cite{aksaray2016q,venkataraman2020tractable,kalagarla2021model}
    and model-based algorithms \cite{kapoor2020model,cohen2023temporal}.} 
    Moreover, RL algorithms have been extended to general MDPs over continuous state spaces for various formal specifications. Deep RL techniques are then leveraged in these cases instead of cumbersome tabular controllers \cite{li2017reinforcement,lavaei2020formal,kazemi2020formal,hasanbeig2020deep,cai2021modular,cai2023overcoming}. It is important to note that while RL algorithms provide asymptotical guarantees for the satisfaction of formal tasks, safety violations may occur during the learning process, which can be unacceptable for safety-critical systems. To address this issue, safe RL techniques, such as online shields \cite{alshiekh2018safe,cheng2019end,carr2023safe}, can be further leveraged to guarantee safety during exploration.
    
    \subsection{Neural-Network-Based Formal Synthesis}
    Finally, in addition to their ability to handle unknown dynamics, data-driven approaches can also enhance the scalability of optimization-based control synthesis. There is a recent trend in using neural networks to generate plans for dynamic systems directly based on given specification formulas. For instance, Recurrent Neural Networks (RNNs) have been employed to generate control sequences for continuous dynamic systems under STL  specifications \cite{liu2021recurrent,hashimoto2022stl2vec}, and Long Short-Term Memory (LSTM) networks have also been explored \cite{leung2022semi}. Although highly efficient, control plans generated by neural networks may lack formal safety guarantees. \change{To ensure safety during the execution process, one may further leverage  model-based CBFs 
    \cite{liu2023safe}, sandboxing architectures \cite{zhong2021safe,zhong2023towards}  or safety shields \cite{alshiekh2018safe,wu2019shield}   to regulated plans  generated by neural networks such that formal safety guarantees can be obtained.
    More recently, neural networks have also been applied to accelerate the standard model-based synthesis algorithms \cite{liu2024nngtl}.}

    \section{Formal Synthesis for Multi-Agent Systems}\label{sec:6}
    In many applications, autonomous systems are characterized as multi-agent systems (MAS), consisting of numerous coupled local modules. A prime example is the swarm robot, where a team of robots collaborates spatially and temporally to achieve a global task. For such multi-agent systems, the system model $\Sigma$ can be expressed as $\Sigma = \Sigma_1 \otimes \Sigma_2 \otimes \cdots \otimes \Sigma_n$, where each $\Sigma_i$ denotes the dynamics of a local agent, and $\otimes$ specifies how their behaviors are synchronized. Furthermore, the overall specification $\varphi$ may consist of a global specification $\varphi_G$ that needs to be satisfied collaboratively and a set of local specifications $\varphi_i$ that only needs to be achieved by each local agent. 

    \change{In the multi-agent setup, however, all three categories of synthesis approaches discussed above face the same challenges. The first and foremost challenge is the curse of dimensionality issue. 
    In theory, if the decision-maker can always maintain the global state information  and control all agents simultaneously, then  the control synthesis problem for MAS can be tackled in a centralized fashion using the approaches mentioned earlier. However, this centralized approach scales poorly as the number of agents increases. 
    For instant, in the context of abstraction-based synthesis for autonomous systems,  suppose that the system involves five homogeneous agents and each agent can be abstracted as a discrete system with only 100 states.  Then, however, the overall synthesis problem will involve at most one hundred billion states, which makes the synthesis problem computationally infeasible. The same design challenge also arises in data-driven control synthesis. For instant, for RL-based synthesis for formal specifications, the state and action spaces also grow exponentially for MAS. 
    Therefore, more scalable approaches, such as top-down approaches, bottom-up approaches or structural approaches, are needed.   
    Also, in many MAS, the information structures for perceptions and control are naturally distributed, further necessitating the use of local control strategies. Finally, there is also a need for new underlying specification languages to capture the interactive nature of MAS. This section will elaborate on recent progress made to address these challenges.}
    \subsection{Top-Down Approaches for MAS Synthesis}
    In the context of MAS formal synthesis, a top-down approach involves a problem-solving strategy where the overall global specification for the entire team is decomposed into smaller and more manageable sub-tasks for each individual agent or group. This approach aligns well with symbolic formal synthesis as it is closely related to the concept of language decomposition in formal language theory \cite{komenda2012conditional,lin2015distributed}. For instance, in the context of multi-robot task planning, researchers in \cite{chen2011formal} explored the decomposition of a global regular language task into individual robot tasks. Additionally,  works such as \cite{schillinger2018simultaneous,schillinger2018decomposition,meyer2019hierarchical,banks2020multi,fang2022decentralized,guo2023hierarchical,li2023fast} have developed hierarchical approaches for decomposing LTL synthesis problems for MAS. The top-down approach for symbolic synthesis, particularly hierarchical control synthesis, has also been extensively studied in the context of supervisory control for discrete-event systems with regular language specifications \cite{wong1996hierarchical,schmidt2010maximally,goorden2019structuring}.

    Top-down approaches for MAS synthesis have also been explored in the context of abstraction-free synthesis for continuous systems under STL   specifications. For example, \cite{charitidou2021signal} presented an approach to decompose STL specifications into local agents using convex optimization techniques. In \cite{leahy2022fast}, Satisfiability Modulo Theories (SMT) is employed to decompose a fragment of STL specifications known as Capability Temporal Logic (CaTL). Furthermore, in \cite{yu2021hierarchical}, a hierarchical control strategy was developed for uncertain discrete-time nonlinear systems with STL specifications.

    \subsection{Bottom-Up Approaches for MAS Synthesis}
    In contrast to the top-down approach, a bottom-up approach involves initiating the synthesis process from the perspective of individual agents, gradually building up to the overall system behavior. A commonly employed bottom-up technique is incremental controller synthesis \cite{hill2008incremental,ulusoy2014incremental}, which explores the independence between each local module. Another powerful approach for bottom-up synthesis is the compositional reasoning \cite{alur2018compositional,kulkarni2018compositional,schuppe2020multi,liu2022compositional,ghasemi2022decentralized}. Specifically, by making assumptions and ensuring guarantees for the interactions with other agents, one can synthesize controllers locally without explicitly enumerating the product state space, while ensuring the satisfaction of the global task. 
    Such approach is also referred to as contract-based synthesis \cite{ghasemi2022decentralized} or interface-based synthesis \cite{leduc2005hierarchical,hill2010multi}. 
    
    \subsection{Structural Properties for MAS Synthesis}
    When dealing with a set of homogeneous agents operating in the same environment, inherent similarities among the agents provide structural properties that can potentially reduce synthesis complexity, particularly in terms of the number of agents involved. For instance, the concept of permutation symmetry was employed in \cite{nilsson2019control} for formal control synthesis of large-scale systems under counting constraints. When the dynamic of each local agent is represented as a symbolic model, the overall MAS can be more efficiently captured as a Petri net model without explicitly enumerating the product state space. In the context of multi-robot planning, several works have concentrated on exploring the structural properties of Petri nets to efficiently synthesize control strategies \cite{mahulea2017robot, kloetzer2020path, mahulea2020path, shi2022path, lv2023optimal}. For example, in \cite{lv2023optimal}, it was demonstrated that using a compact representation of the Petri net state space called Basic Reachability Graph (BRG), the synthesis complexity for a particular type of LTL task can be significantly reduced. 
    
    \subsection{Communication-Aware Synthesis in MAS}
    In the context of formal methods for multi-agent systems, a noteworthy feature distinguishing it from the single-agent setting is the presence of an information structure. Each agent may possess its unique information, shared through a communication network with its neighbors within a certain range. In this context, the synthesis task involves not only finding local control strategies for each agent but also planning their communication protocols. This scenario is termed communication-aware synthesis, presenting additional challenges due to potential information asymmetry among agents. For example, in the context of multi-robot planning, a communication-aware planning framework was introduced for STL specifications \cite{liu2020distributed}. Assuming that agents can only communicate when they are close to each other, control synthesis algorithms have been developed for intermittent communication scenarios, catering to both LTL specifications \cite{guo2018multirobot, kantaros2019temporal} and MTL specifications \cite{wang2022decentralized, xu2023controller}. 
    
    \subsection{MAS Related Formal Specifications}
    Standard formal specifications such as LTL and STL were originally developed for single-agent systems, lacking an explicit consideration of the multi-agent context. In scenarios with a large number of agents, several new semantics for temporal logic specifications have been proposed to effectively capture features specific to multi-agent systems. For instance, in \cite{sahin2019multirobot}, Counting Linear Temporal Logic (cLTL) was introduced to express requirements related to the number of agents simultaneously performing a given task. \change{Similarly,} \cite{cardona2023temporal} extended MTL to the swarm robot setting by incorporating considerations for the maximum number of sub-swarms. Additionally, Swarm Signal Temporal Logic (SwarmSTL) was introduced in \cite{yan2019swarm, yan2022distributed} to address monitoring and control of MAS with real-valued trajectories. Furthermore, the research in \cite{xu2019graph, djeumou2020probabilistic} investigated the inference and control synthesis problems for swarm MAS with spatial-temporal properties, employing the notion of Graph Temporal Logic (GTL).

     \section{Recent Advances and Challenges}\label{sec:7}
    The preceding sections have provided an overview of fundamental aspects in formal control synthesis for safety-critical autonomous systems. However, as autonomous systems operate in complex real-world environments, there are more additional design challenges to consider, in particular, with the rapid developments of network  and AI technologies. In this concluding section, we delve into related research topics, recent advances and research challenges in the field.

    \subsection{Robustness of Controllers}
    In formal control synthesis, most works that provide formal safety guarantees rely on a reliable system model. However, when the behavior of the agent deviates from the nominal model, the safety guarantee may no longer hold, or the performance may change significantly. Therefore, it is crucial to investigate the robustness of the synthesized controller \cite{majumdar2011robust,rungger2015notion,meira2023tolerance,zhang2023robustification}. For instance, the robust semantics of STL  has provided a useful way to quantify the robustness of satisfaction when spatial values change. More recently, temporal robustness has gained attention  capturing how robust the  performance of the controller is when the agent deviates from the nominal trajectory in time \cite{sahin2019multirobot,rodionova2022temporal,rodionova2022combined}. This temporal robustness issue is particularly important for MAS, where agents often operate asynchronously without a global clock \cite{lindemann2022temporal,yu2023efficient,wang2024synthesis}. Additionally, in the formal synthesis of MAS, quantifying the robustness of the entire system concerning the failure of each agent is also an important topic \cite{huang2022failure,huang2023synthesis,zhao2023failure,kalluraya2023resilient}.  
    
    \subsection{Security-Aware Formal Synthesis}
    Existing works on formal control synthesis have primarily focused on safety concerning physical dynamics or correctness regarding logical behaviors. However, with the development of networked control systems, autonomous systems naturally operate in information-rich environments where communications and information transmissions are unavoidable. Therefore, information security has become an increasingly critical issue to consider in formal synthesis, as it is closely related to the safe operation of the system \cite{yin2020approximate,liu2022secure}. 
    Recent efforts have concentrated on synthesizing controllers that ensure the security of the information-flow generated by the system preventing sensitive information from leaking to the outside under the notions of opacity \cite{yang2020secure,xie2021secure,yu2022security,shi2023security} and differential privacy \cite{jones2019towards,xu2020differentially,chen2023differential}. This information-flow security synthesis problem has also been addressed more recently within the general framework of controller synthesis for hyper-properties \cite{clarkson2010hyperproperties,finkbeiner2015algorithms,bonakdarpour2020controller,wang2020hyperproperties,bonnah2023motion,zhao2024unified}. Additionally, many works are investigating the formal synthesis problem for controllers that remain robust against attackers attempting to actively hack into the control or observation channels \cite{niu2020optimal,yao2020attack,kulkarni2021qualitative,ma2021optimal,ma2022resilient,fu2022almost,udupa2022synthesizing,yao2024sensor}.
    
    \subsection{Synthesis under Unstructured Environments}
    From an application perspective, although safety guarantees can be provided, many existing formal synthesis techniques are currently applicable only to structured and ideal environments such as indoor warehouses. Extending formal control synthesis to autonomous systems operating within complex and unstructured environments introduces significant challenges. The primary challenge lies in dealing with uncertain and unknown environments, necessitating the use of sophisticated perception and navigation tools. For instance, autonomous robots in unknown environments usually require to leverage SLAM techniques to build maps and plan concurrently \cite{cristofalo2017localization,kantaros2020reactive,kantaros2022perception,tian2022two,li2022online,kalluraya2023multi,zhou2023local,zhou2023vision}. 
    Due to the use of unverified perception modules, to ensure safety,   one may further need to appropriately quantify perception errors in the presence of rich but uncertain information \cite{li2023refining,puasuareanu2023closed,yang2023safe}.
    
    \subsection{AI-Enabled Formal Synthesis}
    Finally, the development of artificial intelligence (AI) has significantly expanded the potential scope of formal control synthesis for autonomous systems. Deep neural networks, for example, enable autonomous systems equipped with powerful perception modules to classify information in unstructured environments. Moreover, recent developments in large-language models (LLM) offer a more feasible way to make high-level decisions in complex environments \cite{vaswani2017attention,devlin2018bert}. Although AI techniques have significantly enhanced the capabilities of autonomous systems and typically scale very well, the fundamental challenge with AI-enabled approaches is the lack of formal guarantees. Recently, there has been a growing number of works aiming to develop frameworks that ensure formal guarantees with the context AI-enabled safety-critical control synthesis, 
    ranging from formally certifying AI-based control or perception modules     \cite{xiang2018output,huang2020survey,ivanov2021compositional}   
    to quantify uncertainty for AI-based plans to ensure formal guarantees \cite{chen2021reactive,qin2022statistical,calinescu2022discrete,kumar2023conformal,ren2023robots,wang2023conformal,crouse2023formally,lindemann2023safe,lindemann2023conformal}. 
    Also, AI techniques have also been applied to  enable formal synthesis for high-dimensional dynamical systems \cite{bansal2021deepreach,onken2022neural,yang2024learning}.
    In general, there directions are still in their early stages and require more efforts.
    
    \section*{Acknowledgements}
    This work was supported by the National Natural Science Foundation of China (62173226, 62061136004, 62173283)

    \bibliography{mybibfile}
    
    
    
    
    
    
    
    
    
    \end{document}